\documentstyle[epsfig, emulateapj5]{aastex}
\received{}
\revised{}
\accepted{}

\shorttitle{Galactic Outflow}
\shortauthors{Fang et al.}

\reversemarginpar
\makeatletter

\newenvironment{figurehere}
{\def\@captype{figure}}
{}
\makeatother
            
\begin{document}
\title{Signature of Galactic Outflows as Absorption-Free Gaps in the
Ly$\alpha$ Forest} 

\author{Taotao Fang\altaffilmark{1}, Abraham
  Loeb\altaffilmark{2},David Tytler\altaffilmark{3}, David
  Kirkman\altaffilmark{3}, and Nao Suzuki\altaffilmark{3}} 

\begin{abstract}
Powerful outflows from star-forming galaxies are expected to push away
the neutral intergalactic medium (IGM) around those galaxies, and
produce absorption-free gaps in the Ly$\alpha$ forest.  We analyze the
abundance of gaps of various sizes in three high resolution spectra of
quasars at $z\sim$ 3 -- 3.5. The gap statistics agrees well
with a model in which galactic halos above a minimum mass scale of
$M_{\rm min}\sim 10^{10}~M_\odot$ produce bubbles with a
characteristic radius of $R_b\sim 0.48 h^{-1}$ Mpc.  Both numbers are
consistent with naive theoretical expectations, where the minimum
galaxy mass reflects the threshold for infall of gas out of a
photo-ionized IGM.  The observed gaps are typically bounded by deep
absorption features as expected from the accumulation of swept-up gas
on the bubble walls.

\end{abstract}

\keywords{large-scale structure of the universe --- intergalactic
  medium --- quasars: absorption lines --- cosmology: theory}

\altaffiltext{1}{Department of Astronomy, University of California,
Berkeley, CA 94720; {\sl Chandra} Fellow} 

\altaffiltext{2}{Astronomy Department, Harvard University, 60 Garden
Street, Cambridge, MA 02138}

\altaffiltext{3}{Center for Astrophysics and Space Sciences,
  University of California, San Diego, MS~0424, La Jolla, CA 92093}

\section{Introduction}

The formation and evolution of galaxies is regulated by their feedback on
the surrounding intergalactic medium (IGM).  Radiative feedback owing to
photo-ionization heating by a cosmic UV background or hydrodynamic feedback
owing to outflows driven by supernovae or quasars, can suppress the infall
of IGM gas onto low-mass dark-matter halos (see, e.g.,
\citealp{ike86,ree86,bab92,efs92,twe96,sca04}).  The feedback introduces a
minimum halo mass above which halos can efficiently accrete gas from the
IGM and host star formation or quasar activity. Infall of gas is
suppressed if the gravitational potential well of a halo is shallow,
explaining why dwarf galaxies have a much lower abundance than expected for
low-mass halos (e.g., \citealp{whi91,kau93}).

In this {\it Letter}, we examine a simple observational method to
reveal the scars left by hydrodynamic outflows from galaxies on the
IGM.  Powerful outflows from galaxies are expected to push away the
neutral intergalactic medium (IGM) around those galaxies, and produce
absorption-free gaps in the Ly$\alpha$ forest towards a background
quasar \citep{the01}.  Studies of the Ly$\alpha$ forest --- the narrow
absorption lines produced by intergalactic neutral hydrogen in quasar
spectra --- had been very successful in mapping large scale structure
in the universe at high redshift (for a review, see
\citealp{rau98}). While the IGM dynamics is shaped by gravity alone on
large scales (see, e.g., \citealp{cen94,her96,mei01,mcd04}), outflows can
play an important role on small scales by introducing gaps in the
Ly$\alpha$ forest that can be identified in high resolution
observations of quasar spectra.

Our model is simple: an outflow from a galaxy evacuates a bubble in the
IGM, leaving behind little or no neutral hydrogen as the swept-away gas
piles up on the bubble surface. Such a bubble would appear as a gap in the
Ly$\alpha$ forest absorption spectrum of a background quasar, i.e., a flat
spectral segment with little or no absorption. The expected distribution of
absorption-free gaps can be calculated from the number density of halos
above the minimum mass, as only those halos are capable of hosting 
supernova or quasar-driven outflows.  For the halo mass range
of interest, $\sim 10^9$--$10^{11}M_\odot$, the bubble radius is expected
to be almost independent of halo mass at a given redshift (Furlanetto \&
Loeb 2003). Hence, our analytic model has only two free parameters: the
minimum halo mass for the production of outflows, $M_{min}$, and the
characteristic bubble radius $R_b$ for halos above that mass.

There have already been some observational reports about the signatures of
outflows around massive starburst galaxies (with $M\gg M_{\rm min}$) at
high redshifts. \citet{sha03} discovered large scale outflows with
velocities up to $\sim 600\,\rm km\,s^{-1}$ by studying $\sim 10^3$ Lyman
Break Galaxies (LBGs) at $z \sim 3$. Using a large sample of quasar spectra
which pass by foreground LBGs, \citet{ade03} identified the lack of neutral
hydrogen within $\sim 0.5h^{-1}$ comoving Mpc around LBGs. Recent
observations of high redshift, submillimeter galaxies also indicated the
existence of large scale outflows (see, \citealp{gen04} and reference
therein).  The properties of active bubbles around starburst galaxies may
be different than those of mature bubbles around more common galaxies for
which the wind material had sufficient time to accumulate on the bubble
walls and decelerate after evacuating the bubble interior. In
simulations, \citet{cro02} and \citet{kol03} also found absorption-free gaps in their
simulated Ly$\alpha$ spectrum caused by galactic winds. Our goal in this
{\it Letter} is to characterize the properties of the typical relic bubbles
around the most common ($M\sim M_{\rm min}$) galaxies at redshift $z\sim
3$.

Our paper is organized as follows. In \S\ref{sec:calc} we briefly
describe our analytic model and derive its basic equations. In
\S\ref{sec:obs} we compare the model with three high-resolution quasar
spectra. Finally, we conclude with a discussion about the assumptions
of our model and the implications of our results. Throughout the
paper, we assume the concordance cosmological model with
$\Omega_M=0.27, \Omega_{\Lambda}=0.73, \Omega_b=0.044, h=0.71,\, \rm
and\, \sigma_8=0.84$ determined by recent {\sl WMAP} data
\citep{spe03}.

\section{Model} \label{sec:calc}

In our simple model, a galactic wind sweeps-up the surrounding
intergalactic medium and eliminates HI absorption within a bubble radius
$R_b$ of order a few hundred kpc (\citealp{ade03,flo03}). The bubble spends
most of the time approaching a fixed comoving distance from its host
galaxy, at which the (decelerating) expansion speed asymptotes to the local
Hubble velocity (see section \SS\ref{sec:disc}).  Such a bubble would produce a gap in the Ly$\alpha$
forest imprinted on the spectrum of a background quasar. The
absorption-free gap should have a high value of the mean transmission $F$
[defined as $\exp\left(-\tau\right)$ where $\tau$ is the optical depth] that is close
to unity. Our goal is to calculate the expected size distribution of such
gaps in the Ly$\alpha$ forest and compare it with observations.

Given a comoving density of halos per unit mass of $dn_{h}(M)/dM$ and the
assumption that the wind from each halo of mass $M$ produces bubble of a
comoving radius of $R_b(M)$, the probability of finding an absorption-free
gap (with $F \sim 1$) of comoving size in the range between $L$ and $L+dL$
per infinitesimal comoving path-length, $dl$, is given by
\begin{equation}
\frac{dP}{dLdl} = \int_{M_{min}} \frac{dn_{h}}{dM} \frac{d\sigma}{dL} dM.
\end{equation} 
Here, $\sigma=\pi b^2$ is the cross-section of the bubble for an impact
parameter $b=\sqrt{R_b^2-\left(L/2\right)^2}$, and $M_{min}$ is the minimum
halo mass into which IGM gas accretes efficiently.  The differential
cross-section for having a bubble segment of length between $L$ and $L+dL$
is then,
\begin{equation}
\frac{d\sigma}{dL} = \left\{
\begin{array}{ll}
\pi L/2 & 0 < L < 2R_b(M);\\
0              & L \geq 2R_b(M)
  \end{array}
\right.
\label{eq:dsigma}
\end{equation} 
The expected number of absorption-free gaps with a size larger than $L$ is
\begin{eqnarray}
\frac{dP}{dl}(\geq L) & = & \int_{M_{min}}^{\infty} \int_{L}^{2R_b(M)}
\frac{dn_{h}}{dM} \frac{d\sigma}{dL^{\prime}} dL^{\prime} dM
\nonumber \\
                      & = & \int_{M_{min}}^{\infty} 
\pi \left(R_b^2-\frac{L^2}{4}\right) \frac{dn_{h}}{dM} dM.
\label{eq:cumul}
\end{eqnarray} 
Since $R_b(M)$ has a rather weak dependence on $M$ across the relevant
range of halo masses (see Fig. 1 in \citet{flo03}), we assume for
simplicity a single value for $R_b$ when fitting observational data at
a given redshift.  We adopt the \citet{pre74} mass function of halos
and use $M_{min}$ as the second adjustable parameter.  The concordance
cosmological model yields the following values for the cumulative
number density: $n_{h}(>M_{min}) = 1, 0.1,$ and $0.01$~$h^3~\rm
comoving~Mpc^{-3}$, for $M_{min} = 1.3\times10^9, 1.2\times10^{10},$ and
$10^{11}\,\rm M_{\odot}$, respectively. Figure~\ref{f1} shows results from equation~(\ref{eq:cumul}) for
the three cases of $M_{min} = 1.3\times 10^9\,\rm M_{\odot}$ (dashed
line), $M_{min} = 1.2\times10^{10}\,\rm M_{\odot}$ (dotted line),
and $M_{min} = 10^{11}\,\rm M_{\odot}$ (solid line).

\section{Observations} \label{sec:obs}

\subsection{Data Analysis}

Next, we apply our analytic method to three high-resolution, high
signal-to-noise ratio ($S/N$) spectra:  Q~1422+231 at $z=3.62$
\citep{rau01}; Q~0130-4021 at $z = 3.023$ and
Q~0741+4741 at $z = 3.21$ (\citealp{kir00,kir05}). These quasar were observed
with the Keck High Resolution Echelle Spectrometer (HIRES), at $S/N$
of $\gtrsim 50 - 70 $. The resolution for Q~1422+231 is $\sim
23.3h^{-1}$ kpc, and for the other two is $\sim 8\rm\ km\ s^{-1}$. We
refer readers to \citet{rau01} and \citet{kir05} for details of
continuum fitting. By fit artificial spectra for 24 QSOs,
\citet{kir05} measured a typical error in continuum level of
1.2\%. The actually error in our sample should be much smaller due to
the higher $S/N$.

We select a portion of the spectrum around $z \sim 3$ to test our
analytic model. Specifically, we consider the observed wavelength band
of $4500 \leq \lambda \leq 4800$ \AA, corresponding to a comoving path
length of $\sim 180\,h^{-1}$ Mpc. For Q~0130-4021 and Q~0741+4741,
only Ly$\alpha$ absorption presents in this wavelength range, while
for Q~1422+231, Ly$\beta$  will cover part of this range (but not
Ly$\gamma$) and will have quite a bit more aborption. We split the
spectrum into segments with a size of
$ 60\,h^{-1}$ Mpc each so that we have a total of 9 segments for the
entire sample. We then normalize each of the 9 segments separately. We
did this by varying the optical depth $\tau$ so that the mean flux
$\bar{F}=0.684$, consistent with the observed value at $z=3$
(\citealp{mcd00,kir05}). We did not vary the normalization with
wavelength. We measure gaps in each segment, and the errors are
estimated  by using the variance between these 9 subsamples.

We directly measure $dP/dl (\geq L)$ from the observed spectrum. The flat
absorption-free gaps are selected by searching for segments with
transmission $F \geq F_{th}$ where $F_{th}=0.99$ is our fiducial
threshold. Other choices of the
threshold transmission may be considered plausible, since \citet{ade03}
found that the mean transmitted flux is $> 0.8$ within 0.5 $h^{-1}$
comoving Mpc of the LBGs and possibly higher close to the
LBGs. Note that since LBGs are selected by having ongoing starburst
activity, their active wind may add some absorption by outflowing material
inside their bubbles.  Other galaxies are most likely to be surrounded by
empty relic bubbles bounded by a thin shell of swept-up gas, long after
their powerful starburst activity has ended. However, a low value of
$F_{th}$ is very likely to mix the outflow-driven bubbles discussed here
with high-transmission gaps caused by gravitationally-induced voids (which
are underdense but not completely empty). We have chosen the threshold of
0.99 based on a detailed analysis of the results of a numerical
simulation from a Particle-Mesh (PM) code involves gravity only (see. e.g.,
\citealp{mei03,whi99}). We found that the PM simulation provides the
same distribution of gaps as the data up to a threshold of $\sim$0.97,
and the data shows a clear over-abundance of gaps at a threshold of
0.99: at $F_{th}=0.99$ the gravity-induced gaps accounts for less than
$\sim$ 10\% of the total gaps found in the observation data.

In principle we could have used the raw spectrum, the
noise in each pixel generates artificial gaps with a width of just a few
pixels. Instead, we use the fitted spectrum, which smoothes out these
random pixel noise. Using high-order statistic method, \citet{fwh04}
found that the coherence scale length of spatial correlation in Keck
data must be smaller than 0.3$h^{-1}$ Mpc, which roughly corresponds
to 12 pixels in the spectrum. So the raw spectrum is
smoothed by the boxcar method with a width of 12 pixels to ensure
that we can smooth Poisson noise in each pixel but preserve the large
scale features. The 1-$\sigma$ error bars are obtained directly from
Poisson statistics.

We fit the observed distribution of gap sizes using
equation~(\ref{eq:cumul}), with two free parameters: $\rm M_{min}$ and
$R_b$. We use a least-$\chi^2$ analysis with
\begin{equation}
\chi^2 \equiv \sum_i \left[y(M_{min}, R_b) - y_i\right]^2/\sigma_i^2.
\end{equation} 
Here $y(M_{min}, R)$ is the expected value of $dP/dl$ based on
equation~(\ref{eq:cumul}), $y_i$ is the observed value of $dP/dl$, and
$\sigma_i$ is the 1-$\sigma$ error for each data point. The $\chi^2$ fit
yields the best-fit parameters of $\rm M_{min}=1.2 \times10^{10}\,M_{\odot}$
and $R_b=480\,h^{-1}$ comoving kpc, with a $\chi^2$ value of 1.2 per
degree-of-freedom. The model fits the observation data well between 0.1 and 1 $h^{-1}$ Mpc\footnote{The correlation function
of galaxy halos of $\sim 10^{10}\ M_\odot$ at $z\sim 3$ implies that the chance
for overlap of bubbles with a radius $480\,h^{-1}$ comoving kpc around
these galaxies, is small in our data set.}, as indicated by the red solid
line in Figure~\ref{f1}. 

\begin{figurehere}
\begin{center}
\resizebox{2.5in}{!}{\includegraphics[angle=90]{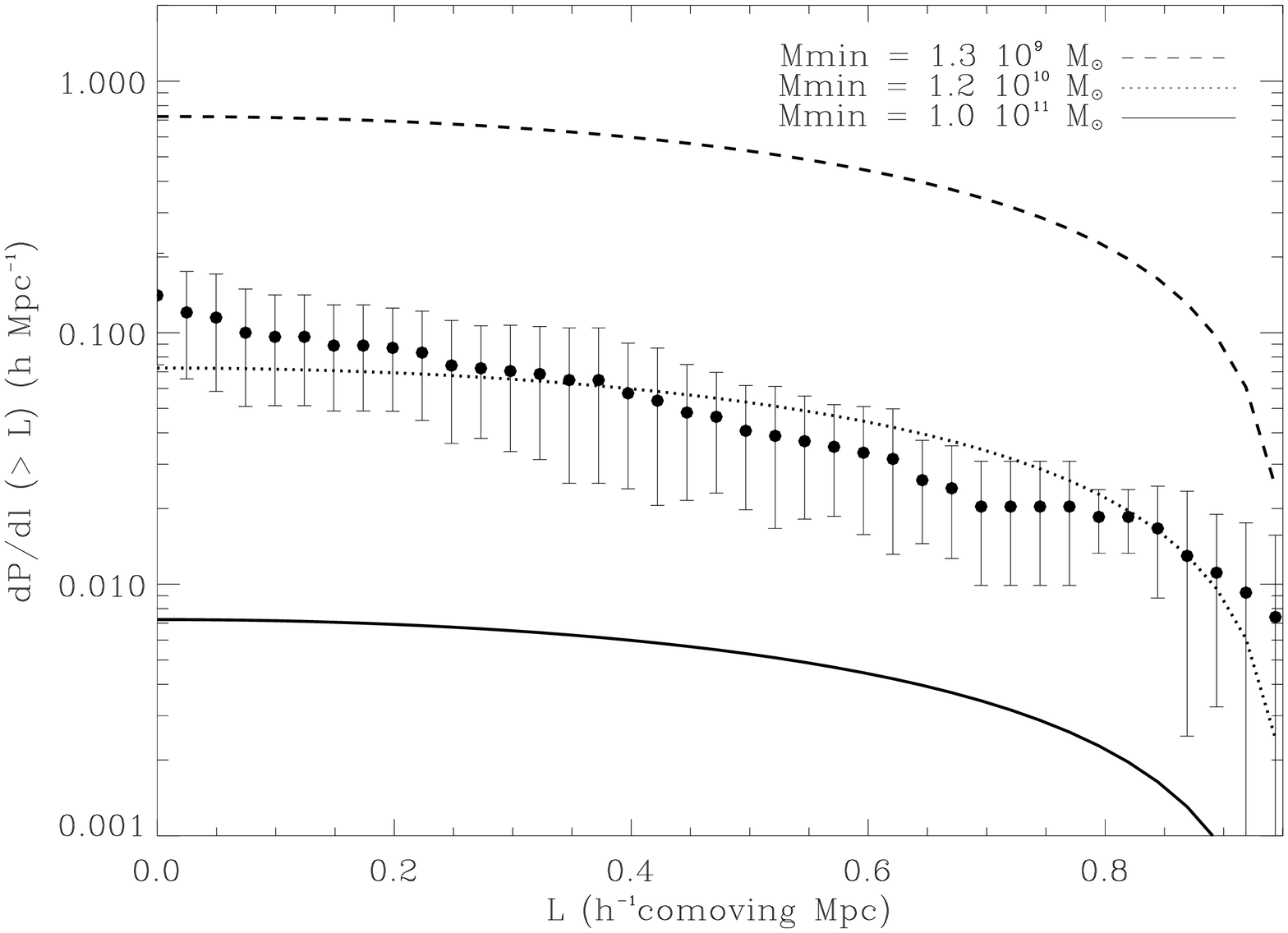}}
\end{center}
\caption{Cumulative distribution of absorption-free gaps with     
comoving length $>L$ per comoving $h^{-1}$ Mpc. The three heavy
lines show results from our a $0.48\,h^{-1}$ Mpc and different
value of $M_{min}$. Data is shown as points (extracted with
$F_{th}=0.99$) with 1--$\sigma$ Poisson error bars. The middle
dashed is our best-fit model.}
\label{f1}
\vskip3pt
\end{figurehere}

\subsection{Results}

The best-fit value for the minimum halo mass that is capable of hosting
outflows (due to star formation or quasar activity), is consistent with the
value expected from hydrodynamic simulations and semi-analytic analysis.
The mass of a dark matter halo is related to its circular velocity
$V_{circ}$ and collapse redshift $z_c$ ($\gg 1$),
\begin{equation}
M = 10^{10} M_{\odot} \left(\frac{V_{circ}}{\rm
50\,km\,s^{-1}}\right)^3 \left(\frac{1+z_c}{4}\right)^{-\frac{3}{2}}.
\label{eq:minm}
\end{equation} 
The infall of gas onto dwarf galaxies is significantly suppressed
through photoionization heating of the IGM by the cosmic ultraviolet
background radiation after cosmological reionization
(\citealp{ree92,efs92,twe96}).  As soon as the IGM is heated to a
temperature of $1$--$2\times 10^4$ K, its thermal pressure suppresses
gas infall into shallow gravitational potential wells. The minimum
halo mass due to photo-ionization heating corresponds to a virial
temperature of $\sim 10^5$ K. The reason is that photo-ionization
heats the IGM to $\sim 10^4$ K but when the gas virializes into a
halo, its density increases by a factor of $\sim 200$ and so its
temperature rises adiabatically by a factor $\sim 200^{2/3} \sim
35$. Only if this final temperature is below the virial temperature,
will the gas condense by this factor of $\sim 200$. The minimum halo
mass for galaxies is defined as the threshold where $\sim 50$\% of the
baryons that were supposed to condense made it. This gives a minimum
virial temperature of $\sim 10^5$ K. The corresponding minimum
circular velocity of the halo is \citep{bar01},
\begin{equation}
V_{min} = 60 \left(T/10^5\rm
  K\right)^{\frac{1}{2}}\rm \,km\,s^{-1}.
\end{equation} Detailed hydrodynamic simulations (e.g., \citealp{twe95,twe96,nav97,dij04})
confirm this value, showing that at redshift $z=3$ only $\sim 50\%$ of the
available gas accretes onto halos with $V_{circ} \sim 50 \rm\,km\,s^{-1}$,
and the infall of gas onto halos with $V_{circ} \leq
25$--$30\rm\,km\,s^{-1}$ is completely suppressed. This yields a minimum
halos mass of $\sim 10^{10}\,M_{\odot}$ based on
equation~(\ref{eq:minm}), in agreement with the value derived from the
statistics of absorption-free gaps in the Ly$\alpha$ forest.

The typical bubble radius $R_b$ we derived is consistent with both
semi-analytic analysis and other observations. Based on a starburst wind
model, \citet{flo03} showed that at redshift $z \sim 3$ and halos with
masses between $\sim 3\times 10^8 \rm\,M_{\odot}$ and $10^{10}
\rm\,M_{\odot}$, the physical (proper) radius $R_b/(1+z)$ ranges between
$\sim$ 50 to $\sim 100$ kpc (see the upper left panel of their Fig.
1). For halo masses above $10^{10} \rm\,M_{\odot}$, the maximum radius can
be achieved by the wind is nearly constant, $\sim 100$ kpc. Observations
indicate the lack of \ion{H}{1} within a radius of 500 comoving kpc around
LBGs, which translates to $\sim 125$ proper kpc at $z=3$ \citep{ade03},
although the interpretation may be complicated \citep{kol03}.

\section{Discussion} \label{sec:disc}

The gas that is swept-up from the bubble volume by a galactic outflow,
is expected to pile-up in a thin shell on the bubble wall, and
introduce enhanced absorption at both ends of its absorption-free
segment in the Ly$\alpha$ forest \citep{flo03}.  Indeed, the spectrum
of Q~1422+231 indicates that almost all of the flat transmission
segments have deep absorption features at one or both of their
boundaries. Figure~\ref{f2} shows four examples of
absorption-free segments with enhanced absorption at both ends. In all
panels, the upper dotted horizontal line show the flux threshold
$F_{th}=0.99$, and the lower horizontal line indicates the mean
transmitted flux of $\bar{F}=0.684$ at $z\sim3$ (bottom). We note that
in analogy with supernova remnants \citep{ch92}, the shell of the
swept-up material may fragment into clumps due to convective or
Rayleigh-Taylor instability. In such a case, a flat (absorption-free)
segment would be bounded by a narrow, deep absorption feature only if
the line-of-sight happens to cross a clump of gas.  We find that out
of 23 absorption-free segments with $L > 0.1h^{-1}$ Mpc, 6 are bounded
at both ends by deep absorption features extending under $\bar{F}$, 12
have such a feature at one end, and only 5 are not bounded at all by a
flux depression under $\bar{F}$. There is tentative evidence for
clumpiness of the bubble walls, although the last class of segments
may also receive a contribution from ionization (Str\"omgren) spheres
around bright galaxies.

\begin{figurehere}
\begin{center}
\resizebox{2.9in}{!}{\includegraphics{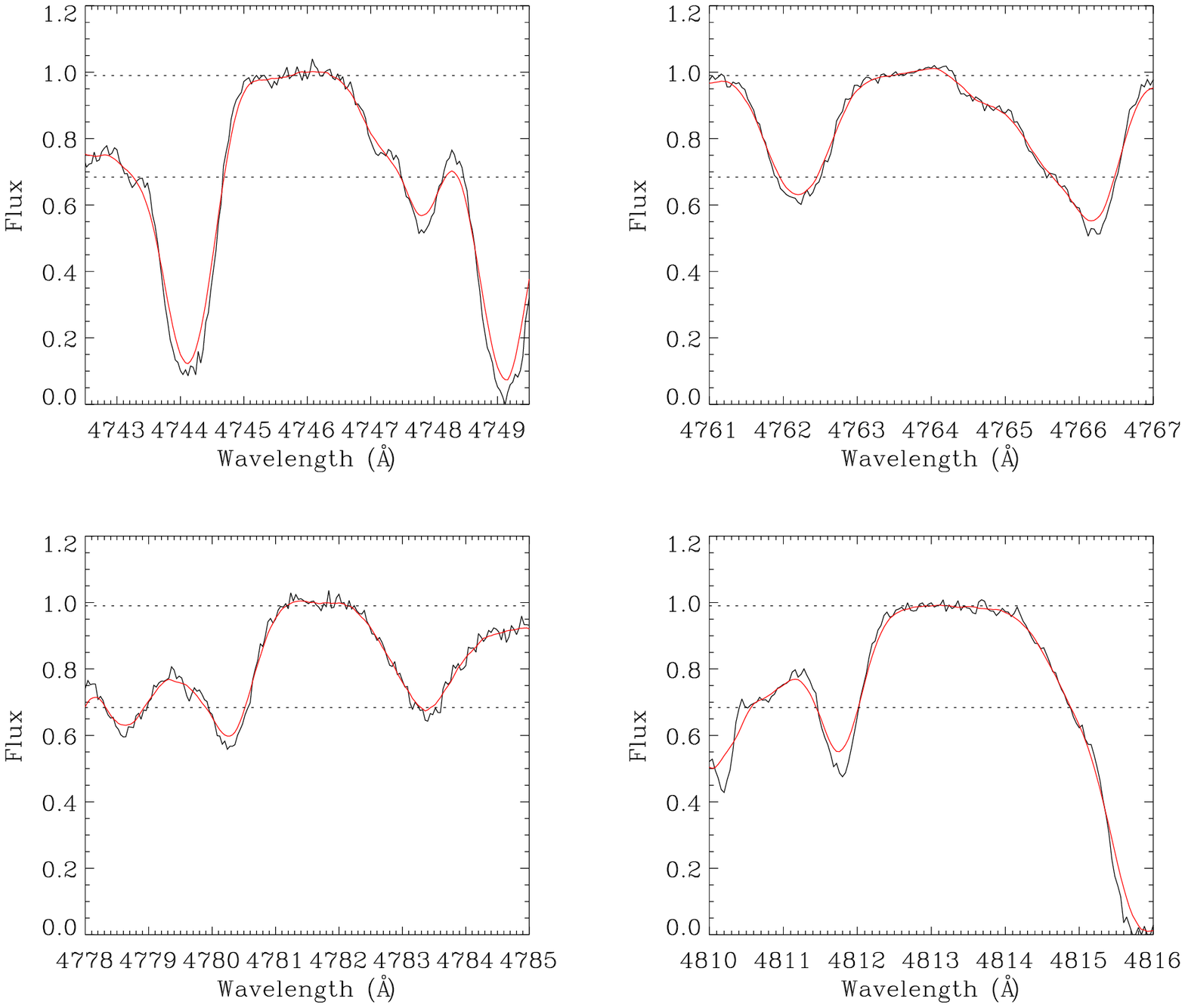}}
\end{center}
\caption{Four examples of absorption-free segments in the         
spectral interval between 4740\AA\,and 4820\AA\,of Q~1422+231. The
raw data is indicated by the black fluctuating lines and the
smoothed spectra are delineated by the red smooth curves. In each
panel, the top horizontal dotted line marks the threshold
$F_{th}=0.99$ and the bottom horizontal line indicates the {\it
mean} transmitted flux at $z\sim3$. All four examples show
enhanced absorption at the two edges of the central flat
(absorption-free) segment, as expected from the pile-up of
hydrogen on the surface of the associated bubbles. Some
absorption-free segments (not shown here) are not bounded by deep
absorption features at both ends, indicating clumpiness of the
bubble material.}
\label{f2}
\vskip3pt
\end{figurehere}

Our analysis assumed that the bubble size can be translated to the spectral
width of the absorption-free gaps for a pure Hubble flow, and ignored the
possible effects peculiar velocities. In reality, the spectral
width may be increased due to the bubble expansion relative to the local
Hubble flow. This would tend to increase the inferred value of $R_b$
relative to its true physical size. The local hubble flow at $z \sim 3$ is
$v_h \sim 32\,\rm km\,s^{-1}$ for bubbles with a radius of $\sim 100$
proper kpc. The upper left panel of Figure 2 in \citet{flo03} shows that
for most halos with halo mass between $10^9$--$10^{10}\,\rm M_{\odot}$
the average bubble expansion velocity is around $v_{exp} \sim 40\,\rm
km\,s^{-1}$. Since the peculiar velocity, $(v_{exp} - v_h)$, is much
smaller than the hubble flow $v_h$, we can safely ignore its contribution
to the broadening of the flat spectral segments. On the other hand,
neutral hydrogen outside the bubble may also have effect in
translating the bubble size into the spectral width of the gap,
depending on whether the bubble is expanding into an infall region
(surrounding an overdense environment) or an outflow region (in a
locally underdense region): the infall can require the true physical bubble
size to be larger (see, e.g., \citealp{kol03}).

In this {\it Letter} we have only attempted to demonstrate the feasibility
of our method at $z\sim 3$.  The statistical significance of our results
can be enhanced through the analysis of additional data sets of
high-resolution quasar spectra. We also need better calibrated data,
with accurate continuum level, and more realistic simulatoins. In the future, our method can be used to
measure the variation of $M_{\rm min}$ and $R_b$ with redshift, and to
gauge the impact of galaxy feedback on estimates of the dark matter
power-spectrum on small scales.

\acknowledgements Quasar spectra were based on data obtained at the
W.M. Keck Observatory, a joint facility of the University of
California, the California Institute of Technology, and NASA. We thank
M.~Rauch and W.~Salgent for providing Keck data of QSO
1422+231, and M.~White for providing PM simulation data. T.~Fang was supported by NASA through the {\sl Chandra}
Postdoctoral Fellowship Award Number PF3-40030 issued by the {\sl
Chandra} X-ray Observatory Center, which is operated by the
Smithsonian Astrophysical Observatory for and on behalf of the NASA
under contract NAS8-39073.  This work was supported in part by NASA
grant NAG 5-13292, and by NSF grant AST-0204514 (for A.L.). DT and DK
were supported in part by NSF grant AST-0098731 and NASA grants
NAG5-13113 and STScI grant HST-AR-10288.01-A.


\begin{thebibliography}{99}

\bibitem[Adelberger et al.(2003)]{ade03} Adelberger, K.~L., Steidel, C.~C.,
Shapley, A.~E., \& Pettini, M.\ 2003, \apj, 584, 45

\bibitem[Babul \& Rees(1992)]{bab92} Babul, A.~\& Rees, 
M.~J.\ 1992, \mnras, 255, 346 

\bibitem[Barkana \& Loeb(2001)]{bar01} Barkana, R. \& Loeb, A.\ 2001,
  Physics Report (astro-ph/0010468)

\bibitem[Cen et al.(1994)]{cen94} Cen, R., Miralda-Escude, J., Ostriker, 
J.~P., \& Rauch, M.\ 1994, \apjl, 437, L9 

\bibitem[Chevalier, Blondin, \& Emmering(1992)]{ch92} Chevalier, R.~A.,
Blondin, J.~M., \& Emmering, R.~T.\ 1992, \apj, 392, 118

\bibitem[Croft et al.(2002)]{cro02} Croft, R.~A.~C., 
Hernquist, L., Springel, V., Westover, M., \& White, M.\ 2002, \apj, 580, 
634 

\bibitem[Dijkstra et al.(2004)]{dij04} 
Dijkstra, M., Haiman, Z., Rees, M.~J., \& Weinberg, D.~H.\ 2004, \apj, 601, 
666 

\bibitem[Efstathiou(1992)]{efs92} Efstathiou, G.~\ 1992, \mnras, 256,
  43 

\bibitem[Fang \& White(2004)]{fwh04} Fang, T., \& White, M.\ 
2004, \apjl, 606, L9 

\bibitem[Furlanetto \& Loeb(2003)]{flo03} Furlanetto, S.~R.~\& Loeb,
  A.\ 2003, \apj, 588, 18  

\bibitem[Genzel et al.(2004)]{gen04} Genzel, R. et al.\ 2004, to
  appear in the proceedings of the Venice conference "Multiwavelength
  Mapping of Galaxy Formation and Evolution" (astro-ph/0403183)

\bibitem[Hernquist et al.(1996)]{her96} 
Hernquist, L., Katz, N., Weinberg, D.~H., \& Miralda-Escude, J.\ 1996, \apjl, 457, 
L51 

\bibitem[Ikeuchi(1986)]{ike86} Ikeuchi, S.\ 1986, \apss, 118, 
509 

\bibitem[Kauffmann, White, \& Guiderdoni(1993)]{kau93} 
Kauffmann, G., White, S.~D.~M., \& Guiderdoni, B.\ 1993, \mnras, 264, 201 

\bibitem[Kirkman et al.(2000)]{kir00} Kirkman, D., Tytler, 
D., Burles, S., Lubin, D., \& O'Meara, J.~M.\ 2000, \apj, 529, 655

\bibitem[Kirkman et al.(2005)]{kir05} Kirkman, D., et al.\ 2005,
  \mnras, accepted (astro-ph/0504391)

\bibitem[Kollmeier et al.(2003)]{kol03} Kollmeier,
J.~A., Weinberg, D.~H., Dav{\' e}, R., \& Katz, N.\ 2003, \apj, 594, 75

\bibitem[Meiksin \& White(2003)]{mei03} Meiksin, A., \& White, M.\ 2003,
  \mnras, 342, 1205 

\bibitem[Meiksin \& White(2001)]{mei01} Meiksin, A., \& 
White, M.\ 2001, \mnras, 324, 141

\bibitem[McDonald et al.(2000)]{mcd00} McDonald, P.,  Miralda-Escud{\'
e}, J., Rauch, M., Sargent, W.~L.~W., Barlow, T.~A., Cen,  R., \&
Ostriker, J.~P.\ 2000, \apj, 543, 1

\bibitem[McDonald et al.(2004)]{mcd04} McDonald, P., et al.\ 2004,
  \apj, submitted (astro-ph/0405013)

\bibitem[Navarro \& Steinmetz(1997)]{nav97} Navarro, J.~F.~\& 
Steinmetz, M.\ 1997, \apj, 478, 13

\bibitem[Peebles(1993)]{pee93} Peebles, P.~J.~E.\ 1993, 
Princeton Series in Physics, Princeton, NJ: Princeton University Press, 
|c1993,  

\bibitem[Press \& Schechter(1974)]{pre74} Press, W.~H.~\& 
Schechter, P.\ 1974, \apj, 187, 425 

\bibitem[Rauch(1998)]{rau98} Rauch, M.\ 1998, \araa, 36, 267 

\bibitem[Rauch et al.(2001)]{rau01} Rauch, M., Sargent, 
W.~L.~W., Barlow, T.~A., \& Carswell, R.~F.\ 2001, \apj, 562, 76

\bibitem[Rees(1986)]{ree86} Rees, M.~J.\ 1986, \mnras, 218, 
25P 

\bibitem[Rees(1992)]{ree92} Rees, M.~J.\ 1992, \mnras, 218, 25 

\bibitem[Scannapieco \& Oh(2004)]{sca04} Scannapieco, E.~\& Oh, S.~P.\
  2004, \apj, accepted (astro-ph/0401087)

\bibitem[Shapley et al.(2003)]{sha03} Shapley, A.~E., Steidel, C.~C.,
Pettini, M., \& Adelberger, K.~L.\ 2003, \apj, 588, 65

\bibitem[Spergel et al.(2003)]{spe03} Spergel, D.~N., et
  al.\ 2003, \apjs, 148, 175 

\bibitem[Theuns et al.(2001)]{the01} Theuns, T., Mo, H.~J., 
\& Schaye, J.\ 2001, \mnras, 321, 450

\bibitem[Thoul \& Weinberg(1995)]{twe95} Thoul, A.~A~\& Weinberg,
  D.~H.\ 1995, \apj, 442, 480

\bibitem[Thoul \& Weinberg(1996)]{twe96} Thoul, A.~A~\& Weinberg,
  D.~H.\ 1996, \apj, 465, 608

\bibitem[White \& Frenk(1991)]{whi91} White, S.~D.~M.~\& 
Frenk, C.~S.\ 1991, \apj, 379, 52 

\bibitem[White(1999)]{whi99} White, M.\ 1999, \mnras, 310, 511

\end{thebibliography}
\end{document}